%% file: syncpeer.tex
\documentclass{article}
\pdfoutput=1
\usepackage{enumerate,ifthen,xspace}
\usepackage{graphicx}
\usepackage{subfig}
\usepackage{amsmath,amsfonts,amstext,amssymb}
\makeatletter\@ifclassloaded{llncs}{}{\usepackage{amsthm}}\makeatother
\usepackage{color}
\usepackage{mdwlist,paralist}
\usepackage{aliascnt,hyperref}
\usepackage[hyperref]{cleveref}

\newcommand{\verboseornot}[2]{#1}
\newcommand{\verboseorextremenot}[2]{#1}
\newcommand{\verboseornotorextremenot}[3]{#1}

\newcommand{\var}[1]{\upshape{\texttt{#1}}}
\newcommand{\val}[1]{\upshape{\texttt{#1}}}

\ifx\pgfversion\undefined
  \input{pgfexternal.tex}
\else

  \usepgflibrary{shapes}
  \tikzstyle{vx}=[circle, draw, inner sep=1pt, minimum width=10pt]
  \tikzstyle{ed}=[->,>=stealth]
  \tikzstyle{ed2}=[ed,<->]
  \tikzstyle{mapping}=[gray,opacity=0.8,semithick] % gray!70,opacity=0.8
  \tikzstyle{bgcircle}=[circle,gray,fill=gray!75,fill opacity=0.3,outer sep=6pt]
  \verboseornot{\tikzstyle{scaled}=[]}{\tikzstyle{scaled}=[scale=0.85,transform shape]}

  \ifthenelse{\lengthtest{\pgfversion cm<1.18cm}}{
  }{
    \ifx\regeneratepgf\undefined\pgfrealjobname{\jobname}
    \else\pgfrealjobname{\regeneratepgf}
    \fi
  }

  \input{pics}
\fi

\newcommand{\true}{\ensuremath{\mathrm{true}}\xspace}
\newcommand{\false}{\ensuremath{\mathrm{false}}\xspace}
\newcommand{\wamoti}[1]{\Sigma^{w\kern-.5ptb}_{#1}\setminus\{\id\}}
\newcommand{\PP}{\ensuremath{\mathcal{P}}\xspace}
\newcommand{\QQ}{\mathcal{Q}}
\newcommand{\RR}{\mathcal{R}}
\newcommand{\occam}{Occam\xspace}

\title{Symmetric and Synchronous Communication in Peer-to-Peer Networks}
\author{Andreas Witzel}
\date{}

\input{util}

\input{mathutil}
\input{csp}

\title{Symmetric and Synchronous Communication\\in Peer-to-Peer Networks}

\makeatletter
\@ifclassloaded{llncs}{
  \author{Andreas Witzel\inst{1, 2}}

  \institute{University~of~Amsterdam, Plantage~Muidergracht~24,
    1018TV~Amsterdam
    \and
    CWI, Kruislaan~413, 1098SJ~Amsterdam, The~Netherlands}

  \bibliographystyle{splncs}
}{
  \author{Andreas Witzel\\
    \normalsize University of Amsterdam, Amsterdam,\\
    \normalsize and CWI, Amsterdam, The Netherlands}
  \bibliographystyle{plain}
}
\makeatother

\begin{document}
\maketitle

% Keywords: communicating sequential processes, direct communication,
%           impossibility result, output guards, symmetry
% Topics: models of computation, programming language semantics

\begin{abstract}
  Motivated by distributed implementations of game-theo\-re\-ti\-cal
  algorithms, we study symmetric process systems and the problem of
  attaining common knowledge between
  processes. We formalize our setting by defining a notion of
  peer-to-peer networks\footnote{Please note that we are \emph{not}
    dealing with fashionable incarnations such as file-sharing
    networks, but merely use this name for a mathematical notion of a
    network consisting of directly connected peers ``treated on an equal
    footing'', i.e. not having a client-server structure or otherwise
    pre-determined roles.}
  and appropriate symmetry concepts in
  the context of Communicating Sequential Processes
  (\CSP)~\cite{hoare_communicating_1978}\verboseornot{, due to the
    common knowledge creating effects of its synchronous communication
    primitives}{}. We then prove that \CSP with input and output guards makes common knowledge
  in symmetric peer-to-peer networks possible, but not the restricted version
  which disallows output statements in guards and is commonly
  implemented. Our results extend~\cite{bouge_existence_1988}.
  \verboseornot{}{

    An extended version is available at \url{http://arxiv.org/abs/0710.2284}\enspace.}
\end{abstract}

\section{Introduction}
\label{sec:introduction}

\subsection{Motivation}
\label{sec:motivation}

Our original motivation comes from the distributed implementation of
game-theoretical algorithms (see e.g.~\cite{halpern_computer_2003} for
a discussion of the interface between game theory and distributed
computing). Two important issues in the domain of
game theory have always been knowledge, especially common knowledge,
and symmetry between the players, also called anonymity. We will
describe these issues and the connections to distributed computing in
the following two paragraphs, before we
motivate our choice of process calculus and the overall goal of the
paper.

\paragraph{Common Knowledge and Synchronization.}
The concept of common know\-ledge has been a topic of much research in
distributed computing as well as in game theory. When
do processes or players ``know'' some fact, mutually
know that they know it, mutually know that they mutually know that
they know it, and so on ad infinitum? And how crucial is the
difference between arbitrarily, but finitely deep mutual knowledge and
the limit case of real common knowledge?

In \verboseornot{the area of}{} distributed computing, the classical example showing
that the difference is indeed essential is the scenario of
Coordinated Attack\verboseornot{, first considered
  by}{}~\cite{gray_notes_1978}. The
game-theoretical incarnation of the underlying issue is the Electronic
Mail Game~\cite{rubinstein_electronic_1989,morris_coordination_2002}.
\verboseorextremenot{

  The basic insight of these examples is that two agents that
  communicate through an unreliable channel can never achieve common
  knowledge, and that their behavior under finite mutual knowledge can be
  strikingly different.

  These issues are analyzed in detail in~\cite{fagin_reasoning_1995},
  in particular in a separately published
  part~\cite{halpern_knowledge_1990}, including a
  variant where communication is reliable, but message delivery
  takes an unknown amount of time. Even in that variant, it is shown
  that only finite mutual knowledge can be attained.
}{
  For a discussion of these examples we refer to the cited papers;
  but the bottom line is that, even with reliable communication but
  unknown delivery time, common knowledge is never achieved, and the
  difference to finite mutual knowledge can be
  striking~\cite{halpern_knowledge_1990}.
}

However, in a synchronous communication act, sending and receiving of
a message is, by definition, performed
simultaneously\verboseorextremenot{. In that way, the agents obtain
  not only the pure factual
  information content of a message, but the sender also knows that the
  receiver has received the message, the receiver knows that the
  sender knows that, and so on ad infinitum. The}{, and in that
way the} communicated
information immediately becomes common knowledge.

%It is thus intuitively clear that
Attaining common knowledge and
achieving synchronization between processes are thus closely
related. Furthermore, synchronization is in itself an important
subject, see e.g.~\cite{schneider_synchronization_1982}.

\paragraph{Symmetry and Peer-to-peer Networks.}
In game theory,
\verboseornot{it is traditionally a fundamental assumption
that players are}{players are assumed to be} anonymous and treated on
an equal footing, in the sense that their names do not play a role and no
single player is a~priori distinguished from the
others~\cite{osborne_introduction_2003,moulin_axioms_1988}.

In distributed computing, too, this kind of symmetry between processes
is \verboseornot{often a desideratum. Reasons}{desirable} to avoid a
predetermined assignment of roles to processes
\verboseornot{or a centralized coordinator include}{and improve}
fault tolerance, modularity, and load
balancing~\cite{andrews_concurrent_1991}.

We will consider symmetry on two levels. Firstly, the communication
network used by the processes should be symmetric to some extent in
order not to discriminate single processes a~priori on a
topological level; we will formalize this requirement by
defining peer-to-peer networks. Secondly, processes in symmetric
positions of the network should have equal possibilities of
behavior; this we will formalize in a semantic symmetry requirement
on the possible computations.

\paragraph{Communicating Sequential Processes (\CSP).}
\verboseorextremenot{
  Since we are interested in synchronization and common knowledge, a
  process calculus which supports synchronous communication
  through primitive statements clearly has some appeal. We will focus on one
  of the prime examples of such calculi, namely}{We focus on}
\CSP, introduced in~\cite{hoare_communicating_1978} and revised
in~\cite{hoare_communicating_1985,schneider_concurrent_1999}\verboseornot{.
  It allows synchronous communication by means of deterministic
  statements on the one hand and non-deterministic alternatives on the
  other hand, where the communication statements occur in guards.

}%
{, since it supports synchronous
  communication through primitive statements.}
Furthermore, it has been implemented in various programming languages,
among the best-known of which is \occam~\cite{inmos_ltd_occam_1988}.
We thus have at our disposal a theoretical framework and programming
tools which in principle could give us
synchronization and common knowledge ``for free''.

However, symmetric situations are a reliable source of
impossibility results~\cite{fich_hundreds_2003}. In particular,
the restricted dialect \CSPin which was, for implementation
issues~\cite{buckley_effective_1983}, chosen to be the theoretical foundation
of \occam is provably~\cite{bouge_existence_1988} less expressive than the
general form, called \CSPio. \CSPin has been used throughout the history of
\occam, up to and including its latest variant \occam-$\pi$~\cite{occam-pi}.
This generally tends to be the case for implementations of \CSP,
one notable exception being a very recent
extension~\cite{welch_integrating_2007} of JCSP\footnote{A
Java\texttrademark{} implementation and extension of \CSP.} to \CSPio.
% CTJ (Hilderink et al., "Communicating Java Threads", "Communicating
% Threads for Java", "A Distributed Real-Time Java System Based on
% CSP") seems to have been left in a prototype stage

Some of the resulting restrictions of \CSPin can in
practice be overcome by using helper processes such as
buffers~\cite{jones_guards_1988}.
Our goal therefore is to formalize the concepts mentioned above, extend the
notion of peer-to-peer networks by allowing helper processes, and
examine whether synchronization is feasible in either of these two
dialects of \CSP. We will come to the result that, while the problem can
(straightforwardly) be solved in \CSPio, it is impossible to do so
in \CSPin.
Our setting thus provides an argument in favor of the former's rare and admittedly more complicated implementations,
such as JCSP.

\subsection{Related Work}
\label{sec:related-work}

This paper builds upon~\cite{bouge_existence_1988}, where a semantic
characterization of symmetry for \CSP is given and fundamental
possibility and impossibility results for the problem of
electing a leader in networks of symmetric processes are proved for
various dialects of \CSP. More recently, this has inspired a similar
work on the more expressive
$\pi$-calculus~\cite{palamidessi_comparing_2003}, but the possibility
of adding helper processes is explicitly excluded.

% \verboseornot{
% The last official standard of the \occam programming language is
% defined in~\cite{sgs-thomson_microelectronics_limited_occam_1995},
% which is an extension of a previously published
% version~\cite{inmos_ltd_occam_1988}.

% }{}
There has been research on how to circumvent problems resulting from
the restrictions of \CSPin. However, solutions
are typically concerned only with the factual content of messages and
do not preserve synchronicity and the common knowledge creating effect
of communication, for example by introducing buffer
processes~\cite{jones_guards_1988}.

The same focus on factual information holds for general
research on synchronizing processes with asynchronous
communication. For example, in~\cite{schneider_synchronization_1982} one goal is to
ensure that a writing process knows that no other process is currently
writing; whether this is common knowledge, is not an issue.

The problem of Coordinated Attack has also been studied for
models in which processes run synchronously~\cite{fich_hundreds_2003};
however, the interesting property of \CSP is that processes run
asynchronously, which is more realistic in physically distributed
systems, and synchronize only at communication statements.

Since we focus on the communication mechanisms, the results will
likely carry over to other formalisms with synchronous
communication facilities comparable to those of \CSP.

\subsection{Overview of the Paper}
\label{sec:overview}

In \cref{sec:preliminaries} we give a short description of
\CSP and the dialects that we are interested in, define some basic
concepts from graph theory, and recall the required notions and
results for symmetric electoral systems from~\cite{bouge_existence_1988}.

In \cref{sec:setting-stage} we \verboseornot{set the stage by first
  formally defining}{formally define} the problem of pairwise
synchronization that we will examine\verboseornot{. Subsequently,
  we}{,} give a formalization of peer-to-peer networks which ensures a
certain kind of symmetry on the topological level, and describe in
what ways we want to allow them to be extended by helper
processes. \verboseornot{Finally, we}{We} adapt a concept
from~\cite{bouge_existence_1988} to capture symmetry on the semantic level.

\Cref{sec:results} contains two positive results and the main
negative result saying that pairwise synchronization of
peer-to-peer networks of symmetric processes is not obtainable in
\CSPin, even if we allow extensions through
buffers or similar helper processes.\verboseornot{

\Cref{sec:conclusions} offers a concluding discussion.}{
\Cref{sec:conclusions} concludes.}

\section{Preliminaries}
\label{sec:preliminaries}
\verboseornot{
We briefly review the required concepts and results from the
\CSP calculus in \cref{sec:csp}, from
graph theory in \cref{sec:graphtheory}, and from~\cite{bouge_existence_1988}
in \cref{sec:symmetricelectoralsystems}. For more details
see~\cite{hoare_communicating_1978,plotkin_operational_1983,bouge_existence_1988}.}{
% We briefly review the required concepts and results, for details
% see~\cite{hoare_communicating_1978,plotkin_operational_1983,bouge_existence_1988}.
}

\subsection{CSP}
\label{sec:csp}

A \emph{\CSP process} consists of a sequential program which can use, besides
the usual \emph{local} statements\verboseornot{ (e.g. assignments or
  expression evaluations involving its local variables)}{}, two
\emph{communication} statements:
\verboseornot{\begin{itemize}}{\begin{compactitem}}
\item \snd{P}{message} to send (output) the given message to process $P$;
\item \rcv{P}{variable} to receive (input) a message
  from \verboseornot{process }{}$P$
  \verboseornot{and store it in the given (local) variable.}%
  {into the given local variable.}
\verboseornot{\end{itemize}}{\end{compactitem}}
Communication is \emph{synchronous}, i.e., send and
receive instructions block until their counterpart is available, at
which point the message is transferred and both participating
processes continue execution. Note that the communication partner $P$
is statically defined in the program code.

There are two \emph{control structures} (see
\cref{fig:controlstructures}).
Each guard is a Boolean expression over local variables
(which, if omitted, is taken to be \true), optionally followed by a
communication statement. A guard is \emph{open} if its Boolean
expression evaluates to \true and its communication statement, if
any, can currently be performed. A guard is \emph{closed} if its
Boolean expression evaluates to \false. Note that a guard can thus be
neither open nor closed.

\begin{figure}[htb]
  \verboseornot{}{\vspace{-4mm}}
  \centering
  \subfloat[Non-deterministic selection]{
    \begin{minipage}{.45\textwidth}
      \begin{csp}
        \If{
          \lAlt{$guard_1$}{$command_1$}\;
          \lOr{$guard_2$}{$command_2$}\;
          $\dots$\;
          \lOr{$guard_k$}{$command_k$}
        }\;
      \end{csp}
    \end{minipage}
  }
  \subfloat[Non-deterministic repetition]{
    \begin{minipage}{.45\textwidth}
      \begin{csp}
        \Do{
          \lAlt{$guard_1$}{$command_1$}\;
          \lOr{$guard_2$}{$command_2$}\;
          $\dots$\;
          \lOr{$guard_k$}{$command_k$}
        }\;
      \end{csp}
    \end{minipage}
  }
  \caption{Control structures in \CSP.}
  \label{fig:controlstructures}
\end{figure}

The selection statement \emph{fails} and execution is aborted
if all guards are closed. Otherwise execution is suspended until
there is at least one open guard. Then one of the open guards is
selected non-deterministically, the required communication (if any)
performed, and the associated command executed.
%(possibly a \skp command without effect).

The repetition statement keeps waiting for, selecting, and executing
open guards and their associated commands until all guards are
closed, and then exits normally; i.e., program execution continues
at the next statement.

We will sometimes use the following abbreviation to denote multiple
branches of a control structure (for some finite set $X$):
\verboseornot{\begin{csp}
  \lAlt{$\choice_{x\in X}guard_x$}{$command_x$}\;
\end{csp}}{%
\makebox[17pc]{\begin{csp}
  \lAlt{$\choice_{x\in X}guard_x$}{$command_x$}\;
\end{csp}
}}

Various dialects of \CSP can be distinguished according to what kind
of communication statements are allowed to appear in
guards. Specifically, in \CSPin only input statements are allowed,
and in \CSPio both input and output statements are allowed (within the
same control structure). For technical reasons, \CSPin has been
suggested from the beginning~\cite{hoare_communicating_1978} and is
indeed commonly used for implementations, as mentioned in
\cref{sec:motivation}.

\begin{definition}
  A \emph{communication graph} (or \emph{network}) is a directed graph
  without self-loops. A \emph{process system} (or simply
  \emph{system}) \PP with
  communication graph $G=(V,E)$ is a set of component processes $\{P_v\}_{v\in
    V}$ such that for all $v,w\in V$, if the program run by $P_v$
  (resp. $P_w$) contains an output command to $P_w$ (resp. input
  command from $P_v$) then $(v,w)\in E$. In that case we say that
  $G$ \emph{admits} \PP. We identify vertices $v$ and associated
  processes $P_v$ and use them interchangeably.
\end{definition}

\begin{figure}[htb]
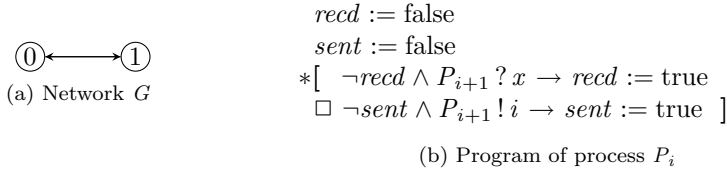

  \verboseornot{}{\vspace{-8mm}}
  \centering
  \subfloat[Network $G$]{
    \qquad
    \beginpgfgraphicnamed{graphic1}%
    \exampleSimpleNetwork%
    \endpgfgraphicnamed%
    \qquad
  }
  \quad
  \subfloat[Program of process $P_i$]{
    \quad
    \begin{minipage}{.6\textwidth}
      \begin{csp}
      $recd\gets\false$\;
      $sent\gets\false$\;
        \Do{
          \lAlt{$\neg recd\wedge\rcv{P_{i+1}}{x}$}{$recd\gets\true$}\;
          \lOr{$\neg sent\wedge\snd{P_{i+1}}{i}$}{$sent\gets\true$}
        }\;
      \end{csp}
    \end{minipage}
    \quad
  }
  \caption[Network and program run by $P_0$ and $P_1$ in
    \cref{ex:csp}.]{Network and program run by $P_0$ and $P_1$ in
    \cref{ex:csp}. Addition of process names here and in all
    further example programs is modulo $2$.}
  \label{fig:syncCSPio}
  \verboseornotorextremenot{}{\vspace{-5mm}}{\vspace{-7mm}}
\end{figure}

\begin{example}
  \label{ex:csp}
  \Cref{fig:syncCSPio} shows a simple network $G$ with the vertex names
  written inside the vertices, and a \CSPio program run by two
  processes which make up a system $\PP:=\{P_0,P_1\}$. Obviously, $G$
  admits $\PP$. The intended behavior is that the processes send each
  other, in non-deterministic order, a message containing their
  respective process name.\verboseornot{
  Since communication is synchronous, it is guaranteed that both processes
  execute each communication statement synchronously at the time when
  the message is transmitted. In a larger context, executing this code
  fragment would have the effect that the participating processes
  synchronize, i.e., wait for each other and jointly perform the
  communication. In terms of knowledge, this fact as well as the
  transmitted message (which can of course be more interesting than
  just the process names) become common knowledge between the processes.}{}
\end{example}

\begin{definition}
  A \emph{state} of a system \PP is the collection of all component
  processes' (local) variables together with their current execution
  positions. A \emph{computation step} is a transition from one state
  to another, involving either one component
  process executing a local statement, or two component processes
  jointly executing a pair of matching (send and receive) communication statements.
  % define active, idle, yielded state if necessary
  The valid computation steps are determined by the state of the
  system.

  A \emph{computation} is a maximal sequence of valid computation
  steps, i.e. a sequence which is not a prefix of any other sequence
  of valid computation steps. A computation%
  \verboseornotorextremenot{\begin{itemize}}{\begin{compactitem}}{}%
  \verboseorextremenot{\item}{} is \emph{properly terminated} if all
  component processes have completed their last instruction,%
  \verboseorextremenot{\item}{} \emph{diverges} if it is infinite, and%
  \verboseorextremenot{\item}{} is in \emph{deadlock} if it is finite
  but not properly terminated.%
  \verboseornotorextremenot{\end{itemize}}{\end{compactitem}}{}%
\end{definition}

\verboseornot{
\begin{example}
  \label{ex:computation}
  \Cref{fig:computation} shows a computation of the system from
  \cref{fig:syncCSPio}. It is finite and both processes reach the end of
  their respective program, so it is properly terminated. Note that
  the exact order in which, for example, the processes get to
  initialize their local variables is non-deterministic, so there are
  other computations with these steps exchanged. Only certain
  restrictions to the order apply, e.g. that the steps within one
  process are ordered corresponding to its program, or that both
  processes must have evaluated the Boolean guards before they can
  participate in the subsequent communication.
\end{example}

\begin{figure}[htb]
  \centering
  \scriptsize
  \begin{align*}
    P_0 &:\text{assign \false to $recd$}\\[-3pt]
    P_1 &:\text{assign \false to $recd$}\\[-3pt]
    P_1 &:\text{assign \false to $sent$}\\[-3pt]
    P_0 &:\text{assign \false to $sent$}\\[-3pt]
    P_1 &:\text{evaluate Boolean guards}\\[-3pt]
    P_0 &:\text{evaluate Boolean guards}\\[-3pt]
    P_0,P_1 &:\text{send $0$ from $P_0$ to $P_1$'s variable $x$}\\[-3pt]
    P_0 &:\text{assign \true to $sent$}\\[-3pt]
    P_0 &:\text{evaluate Boolean guards}\\[-3pt]
    P_1 &:\text{assign \true to $recd$}\\[-3pt]
    P_1 &:\text{evaluate Boolean guards}\\[-3pt]
    P_0,P_1 &:\text{send $1$ from $P_1$ to $P_0$'s variable $x$}\\[-3pt]
    P_1 &:\text{assign \true to $sent$}\\[-3pt]
    P_0 &:\text{assign \true to $recd$}\\[-3pt]
    P_0 &:\text{evaluate Boolean guards and exit repetition}\\[-3pt]
    P_1 &:\text{evaluate Boolean guards and exit repetition}
  \end{align*}
  \vspace{-8mm}
  \caption{A properly terminating computation of the system from \cref{ex:csp}.}
  \label{fig:computation}
\end{figure}
}{}

\subsection{Graph Theory}
\label{sec:graphtheory}

We state some fundamental notions concerning directed finite graphs,
from here on simply referred to as graphs.
\begin{definition}
  Two vertices $a,b\in V$ of a graph $G=(V,E)$ are \emph{strongly
    connected} if there are paths from $a$ to $b$ and from $b$
  to $a$; $G$ is strongly connected if all pairs of vertices are.
  \verboseornot{

  }{}
  Two vertices $a,b\in V$ are \emph{directly connected} if $(a,b)\in
  E$ or $(b,a)\in E$; $G$ is directly connected if all pairs of
  vertices are.
\end{definition}

\begin{definition}
  An \emph{automorphism} of a graph $G=(V,E)$ is a permutation
  $\sigma$ of $V$ such that for all $v,w\in V$,
  \verboseornot{\begin{equation*}}{$}
    (v,w)\in E\text{ implies }(\sigma(v),\sigma(w))\in E
  \verboseornot{\enspace.\end{equation*}}{$.}
  The \emph{automorphism group $\Sigma_G$} of a graph $G$ is the set
  of all automorphisms of $G$.
  The least $p>0$ with $\sigma^p=\id$ is called the \emph{period} of
  $\sigma$, where by $\id$ we denote the identity function
  defined on the domain of whatever function it is compared to.

  The \emph{orbit} of $v\in V$ under $\sigma\in\Sigma_G$ is
  $O_v^\sigma:=\{\sigma^p(v)|p\geq0\}$.
  An automorphism $\sigma$ is \emph{well-balanced} if the orbits of
  all vertices have the same cardinality, or alternatively, if for all
  $p\geq0$,
  \begin{equation*}
    \sigma^p(v)=v\text{ for some $v\in V$}
    \text{ implies }\sigma^p=\id\enspace.
  \end{equation*}
  We will usually consider the (possibly empty) set $\wamoti{G}$ of
  \emph{non-trivial} well-balanced automorphisms of a graph $G$, that
  is those with period greater than~$1$.

  A subset $W\subseteq V$ is called \emph{invariant} under
  $\sigma\in\Sigma_G$ if $\sigma(W)=W$\verboseornot{, i.e. if $W$ is
    an orbit under $\sigma$}{}; it is called invariant under $\Sigma_G$
  if it is invariant under all $\sigma\in\Sigma_G$.
\end{definition}

\begin{figure}[htb]
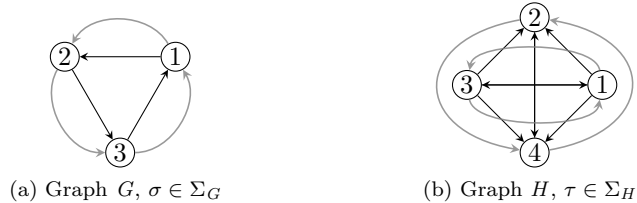

%  \verboseornotorextremenot{}{\vspace{-5mm}}{\vspace{-9mm}}
  \centering
  \hfill
  \subfloat[Graph $G$, $\sigma\in\Sigma_G$]{
    \label{fig:exampleGraph:p2p}
    \qquad
    \beginpgfgraphicnamed{graphic2}%
    \exampleNetworkAndExtension{}{x}%
    \endpgfgraphicnamed%
    \qquad
  }
  \hfill
  \subfloat[Graph $H$, $\tau\in\Sigma_H$]{
    \label{fig:exampleGraph:alsop2p}
    \qquad
    \beginpgfgraphicnamed{graphic3}%
    \exampleGraph%
    \endpgfgraphicnamed%
    \qquad
  }
  \hfill{}
  \caption[Two graphs with non-trivial well-balanced
  automorphisms.]{Two graphs with non-trivial well-balanced
    automorphisms, indicated by gray, bent arrows.}
  \label{fig:exampleGraph}
  \verboseornot{}{\vspace{-5mm}}
\end{figure}

\begin{example}
  \Cref{fig:exampleGraph} shows two graphs $G$ and $H$
  and\verboseornot{}{ well-balanced} automorphisms $\sigma\in\Sigma_G$
  with period $3$ and
  $\tau\in\Sigma_H$ with period $2$.
  \verboseornot{Both are well-balanced since, e.g.,
    $O_1^\tau=O_3^\tau=\{1,3\}$ and $O_2^\tau=O_4^\tau=\{2,4\}$
    all have the same cardinality.}{}
  We have $\Sigma_H=\{\id,\tau\}$, so
  $\{1,3\}$ and $\{2,4\}$ are invariant under $\Sigma_H$.
\end{example}

\subsection{Symmetric Electoral Systems}
\label{sec:symmetricelectoralsystems}

We take over the semantic definition of symmetry
from~\cite{bouge_existence_1988}. As discussed there,
syntactic notions of symmetry are difficult to formalize properly;
requiring that ``all processes run the same program'' does not do the job.
We will skip the formal details since we are not going to use them. The
interested reader is referred to~\cite{bouge_existence_1988}.

\begin{definition}[{adapted from~\cite[Definition~2.2.2]{bouge_existence_1988}}]
  \label{dfn:symmetricsystem}
  A system \PP with communication graph $G=(V,E)$ is \emph{symmetric}
  if for each automorphism $\sigma\in\Sigma_G$ and each computation
  $C$ of \PP, there is a computation $C'$ of \PP in which, for each
  $v\in V$, process $P_{\sigma(v)}$ performs the same steps as $P_v$ in $C$,
  modulo changing via $\sigma$ the process names occurring in the
  computation (e.g. as communication partners).
\end{definition}

The intuitive interpretation of this symmetry notion is as
follows. Any two processes which are not already distinguished by the
communication graph~$G$ itself, i.e. which are related by some
automorphism, must have equal possibilities of behavior.
That is, whatever behavior one process exhibits in some particular
possible execution of the system (i.e., in some computation), the other
process must exhibit in some other possible execution of the system,
localized to its position in the graph by appropriate process
renaming. Taken back to the syntactic level, this can be achieved by
running the same program in both processes, which must not make use of
any externally given distinctive features like, for example, an
ordering of the process names.

\begin{example}
  \label{ex:symmetric}
  The system from \cref{fig:syncCSPio} is symmetric. It is easy to see
  that, \verboseornot{for example, if we swap all $0$s and $1$s in
    the computation shown in \cref{fig:computation}}{if we swap all
    names $0$ and $1$ in any computation of \PP}, we still have a
  computation of \PP.
  Note that programs are allowed to access the process names,
  and indeed they do; however, they do not, for example, use their
  natural order to determine which process sends first.
\end{example}

\begin{example}
  \label{ex:asymmetric}
  On the other hand,\verboseornot{ consider}{} the system $\QQ=\{Q_0,Q_1\}$ 
  \verboseornot{running the program in
    \cref{fig:asymmetric}.
    There is obviously a
    computation where $Q_0$ sends its process name $0$ to $Q_1$; since
    the two vertices of the communication graph are related by an
    automorphism, symmetry would require that there also be a
    computation where $Q_1$ sends its process name $1$ to
    $Q_0$. However, such a computation does not exist due to the use of
    the process name for determining the communication role, so the
    system is not symmetric.
  }{where each $Q_i$ runs the following program is
    not symmetric:
    \small\begin{csp}
      \If{
        \lAlt{$i=0$}{$\snd{Q_{i+1}}{i}$}\;
        \lOr{$i=1$}{$\rcv{Q_{i+1}}{x}$}
      }\;
    \end{csp}\normalsize}
\end{example}
\verboseornot{
\begin{figure}[htb]
  \centering
  \begin{minipage}{.4\textwidth}
    \begin{csp}
      \If{
        \lAlt{$i=0$}{$\snd{Q_{i+1}}{i}$}\;
        \lOr{$i=1$}{$\rcv{Q_{i+1}}{x}$}
      }\;
    \end{csp}
  \end{minipage}  
  \caption[Asymmetric program run by $Q_0$ and $Q_1$ in
    \cref{ex:asymmetric}.]{Asymmetric program run by
      $Q_0$ and $Q_1$ in \cref{ex:asymmetric}.}
  \label{fig:asymmetric}
\end{figure}
}{}

We now recall a classical problem for networks of processes,
\verboseornot{pointed out by~\cite{le_lann_distributed_1977}.}{and
  then restate the impossibility result which our paper builds on.}

\begin{definition}[{from~\cite[Definition~1.2.1]{bouge_existence_1988}}]
  A system \PP is an \emph{electoral system} if
  \verboseorextremenot{\begin{enumerate}[(i)]}{\begin{inparaenum}[(i)]}
  \item all computations of \PP are properly terminating and
  \item each process of \PP has a local variable \var{leader}, and at
    the time of termination all these variables contain the same
    value, namely the name of some process $P\in\PP$.
  \verboseorextremenot{\end{enumerate}}{\end{inparaenum}}
\end{definition}

\verboseornot{
We now restate the impossibility result which our paper builds
on, combining a graph-theoretical characterization with the symmetry
notion and electoral systems.
}{}

\begin{theorem}[{from~\cite[Theorem~3.3.2]{bouge_existence_1988}}]
  \label{thr:bouge:3.3.2}
  Suppose a network $G$ admits some well-balanced automorphism
  $\sigma$ different from $\id$. Then $G$ admits no symmetric electoral
  system in \CSPin.
\end{theorem}

\section{Setting the Stage}
\label{sec:setting-stage}

\subsection{Pairwise Synchronization}
\label{sec:pairw-synchr}

Intuitively, if we look at synchronization as part of a larger system,
a process is able to synchronize with another process
if it can execute an algorithm such that a direct communication (of
any message) between the two processes takes place. This may be
the starting point of some communication protocol to exchange more
information, or simply be taken as an event creating common knowledge
about the processes' current progress of execution.

Communication in \CSP always involves exactly two processes and
facilities for synchronous broadcast do not exist, thus
synchronization is inherently pairwise only. This special case is
still interesting and has been subject to research, see
e.g.~\cite{parikh_communication_1990}.

Focusing on the synchronization algorithm, we want to guarantee that
it allows all pairs of processes to synchronize. To this end,
we require existence of a system where in all computations, all pairs
of processes synchronize. Most probably, in a real system not all pairs of
processes need to synchronize in all executions. However, if one
has an algorithm which in principle allows that, then one could
certainly design a system where they
actually do; and, vice versa, if one has a system which is guaranteed
to synchronize all pairs of processes, then one can
obviously use its algorithms to synchronize any given pair. Therefore
we use the following formal notion.

\begin{definition}
  A system \PP of processes \emph{(pairwise) synchronizes}
  $\QQ\subseteq\PP$ if all computations of \PP are finite and properly
  terminating and contain, for each pair $P_a,P_b\in\QQ$,
  at least one direct communication from $P_a$ to $P_b$ or from $P_b$ to $P_a$.
\end{definition}

\begin{example}
  The system from \cref{fig:syncCSPio} synchronizes
  $\{P_0,P_1\}$.
\end{example}

Note that the program considered so far is not a valid \CSPin
program, since there an output statement appears within a guard. If we
want to restrict ourselves to \CSPin (for example, to
implement the program in \occam), we have to get rid of that
statement. Attempts to simply move it out of the guard fail since the
symmetric situation inevitably leads to a system which may deadlock.

To see this, consider the system $\PP'=\{P_0',P_1'\}$
\verboseornot{with the program shown in \cref{fig:workaround-bad}.}%
{running the following program:

  \begin{minipage}{.6\textwidth}
    \small\begin{csp}
      $recd\gets\false$\;
      $sent\gets\false$\;
      \Do{
        \lAlt{$\neg recd\wedge\rcv{P_{i+1}'}{x}$}{$recd\gets\true$}\;
        \lOr{$\neg sent$}{$\snd{P_{i+1}'}{i}$; $sent\gets\true$}
      }\;
    \end{csp}\normalsize
  \end{minipage}

}
There is no guarantee that not both processes enter the second clause
of the repetition
at the same time and then block forever at the output statement,
waiting for each other to become ready for input.
A standard workaround~\cite{jones_guards_1988} for such cases is to
introduce buffer processes mediating between the main processes, in
our case resulting in the extended system $\RR=\{R_0,R_0',R_1,R_1'\}$%
\verboseornot{
shown in \cref{fig:workaround-buffer}.

\begin{figure}[htb]
%  \verboseornot{}{\vspace{-5mm}}
  \centering
  \begin{minipage}{.6\textwidth}
    \begin{csp}
      $recd\gets\false$\;
      $sent\gets\false$\;
      \Do{
        \lAlt{$\neg recd\wedge\rcv{P_{i+1}'}{x}$}{$recd\gets\true$}\;
        \lOr{$\neg sent$}{$\snd{P_{i+1}'}{i}$; $sent\gets\true$}
      }\;
    \end{csp}
  \end{minipage}
  \caption{Program of process $P_i'$ potentially causing deadlock.}
  \label{fig:workaround-bad}
%  \verboseornot{}{\vspace{-5mm}}
\end{figure}

\begin{figure}[htb]
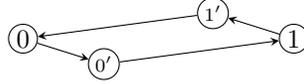

%  \verboseornot{}{\vspace{-5mm}}
  \centering
  \subfloat[Program of main process $R_i$]{
    \begin{minipage}{.55\textwidth}
      \begin{csp}
        $recd\gets\false$\;
        $sent\gets\false$\;
        \Do{
          \lAlt{$\neg recd\wedge\rcv{R_{i+1}'}{x}$}{$recd\gets\true$}\;
          \lOr{$\neg sent$}{\snd{R_i'}{i}; $sent\gets\true$}
        }\;
      \end{csp}
    \end{minipage}
  }
  \subfloat[Program of buffer process $R_i'$]{
    \qquad\quad
    \begin{minipage}{.3\textwidth}
      \begin{csp}
        $\rcv{R_i}{y}$\;
        $\snd{R_{i+1}}{y}$\;
      \end{csp}
    \end{minipage}
  }\\
  \subfloat[Underlying communication network]{
    \qquad\quad
    \beginpgfgraphicnamed{graphic4}%
    \exampleSimpleNetworkBuf%
    \endpgfgraphicnamed%
    \qquad\quad
  }
  \caption[Extended system with main and buffer processes.]{Extended
    system with main processes $R_0$ and $R_1$ and
    buffer processes $R_0'$ and $R_1'$ together with the underlying
    communication network.}
  \label{fig:workaround-buffer}
  \verboseornot{}{\vspace{-5mm}}
\end{figure}
}{:\vspace{2mm}

    \begin{minipage}[b]{.55\textwidth}
      \small\begin{csp}
        $recd\gets\false$\;
        $sent\gets\false$\;
        \Do{
          \lAlt{$\neg recd\wedge\rcv{R_{i+1}'}{x}$}{$recd\gets\true$}\;
          \lOr{$\neg sent$}{\snd{R_i'}{i}; $sent\gets\true$}
        }\;
      \end{csp}\normalsize
      (program of main process $R_i$)
    \end{minipage}
    \begin{minipage}[b]{.4\textwidth}
      \small\begin{csp}
        $\rcv{R_i}{y}$\;
        $\snd{R_{i+1}}{y}$\;
      \end{csp}\normalsize
      (program of buffer process $R_i'$)
    \end{minipage}
}

While the actual data transmitted
between the main processes remains the same, this system obviously
cannot synchronize $\{R_0,R_1\}$,
since there is
\verboseornot{not even a direct link in the communication network.}%
{no direct communication between them.}
This removes the synchronizing and common knowledge
creating effects of communication. \verboseornot{Even though a buffer
  might notify its main process when its message is delivered, then
  notify the communication partner about the notification, and so on,
  synchronicity is not restored and mutual knowledge only}{Mutual knowledge
  can only be} achieved to a finite (if arbitrarily high) level, as
discussed in \cref{sec:motivation}.

The obvious question now is: Is it possible to
change the program or use buffer or other helper processes in more
complicated and smarter ways to negotiate between the main processes
and aid them in establishing direct communications?

To attack this question, in the following \cref{sec:peer-to-peer-networks} we
will formalize the kind of communication networks we are
interested in and define how they may be extended in order to allow
for helper processes without affecting the symmetry inherent in the
original network.

\subsection{Peer-to-peer networks}
\label{sec:peer-to-peer-networks}
%\verboseornot{
The idea of peer-to-peer networks is to have nodes which can
communicate with each other directly and on an equal footing,
i.e. there is no predetermined client/server architecture or central
authority coordinating the communication. We first formalize the
topological prerequisites for this, and then adapt the semantic
symmetry requirement to our setting.
%}{}

\begin{definition}
  \label{dfn:p2p}
  A \emph{peer-to-peer network} is a communication graph $G=(V,E)$
  with at least two vertices (also called nodes) such that
  \verboseornotorextremenot{\begin{enumerate}[(i)]}{\begin{compactenum}[(i)]}{\begin{inparaenum}[(i)]}
  \item $G$ is strongly connected,
  \item $G$ is directly connected, %for any two vertices $a,b\in V$, $(a,b)\in E$ or $(b,a)\in E$,
    and
  \item we have $\wamoti{G}\neq\emptyset$.
  \verboseornotorextremenot{\end{enumerate}}{\end{compactenum}}{\end{inparaenum}}
\end{definition}

In this definition,
\begin{inparaenum}[(i)]
  \item says that each node has the possibility to contact (at least
    indirectly) any other node, reflecting the fact that there are no
    predetermined\verboseornot{ client/server}{} roles;
  \item ensures that all pairs of nodes have a direct connection at
    least in one direction, without which pairwise synchronization by
    definition would be impossible; and
  \item requires a kind of symmetry in the network.
\end{inparaenum}
This last item is implied by the more intuitive requirement that
there be some $\sigma\in\Sigma_G$ with only one orbit, i.e. an
automorphism relating all nodes to each other and thus making sure
that they are topologically on an equal footing. The requirement we
use is less restrictive and suffices for our purposes.

\begin{example}
  \label{ex:p2p}
  See \cref{fig:exampleGraph} for two
  examples of peer-to-peer networks.
\end{example}

We will consider extensions \verboseornot{of a
  peer-to-peer network which we will consider in order to
  allow}{allowing} for helper processes while preserving
the symmetry inherent in the network. \verboseornot{The intuitive
  background for this kind of extensions is that we}{We} view the
peers, i.e. the nodes of the original network, as processors each
running a main process, while the added nodes can be thought of as
helper processes running on the same processor as their respective
main process.\verboseornot{ Communication connections between
  processors are physically given, while inside a processor they can
  be created as necessary.}{}

\begin{definition}
  \label{dfn:p2pext}
  Let $G=(V,E)$ be a peer-to-peer network, then
  $G'=(V',E')$ is a \emph{symmetry-preserving extension of $G$}
  iff there is a collection $\{S_v\}_{v\in V}$ partitioning $V'$ such that
  \verboseornot{\begin{enumerate}[(i)]}{\begin{compactenum}[(i)]}
  \item\label{dfn:p2pext:vinSv} for all $v\in V$, we have $v\in S_v$;
  \item\label{dfn:p2pext:Svconn} all $v\in V$\verboseornot{ and}{,} $v'\in
    S_v\setminus\{v\}$ are strongly connected
    (possibly via nodes $\not\in S_v$);
  \item\label{dfn:p2pext:interSv} for all $v,w\in V$,
    $E'\cap(S_v\times S_w)\neq\emptyset$
    iff $(v,w)\in E$;
  \item\label{dfn:p2pext:iotaS} there is, for each
    $\sigma\in\Sigma_G$, an automorphism $\iota_\sigma\in\Sigma_{G'}$
  extending $\sigma$ such that $\iota_\sigma(S_v)=S_{\sigma(v)}$ for
  all $v\in V$.
  \verboseornot{\end{enumerate}}{\end{compactenum}}
\end{definition}

\begin{remark}
  In general, the collection $\{S_v\}_{v\in V}$ may not be
  unique. When we refer to it, we implicitly fix an arbitrary one.
\end{remark}

\noindent Intuitively, these requirements are justified as follows:
\begin{enumerate}[(i)]
\item Each $S_v$ can be seen as the collection of processes running on the
  processor at vertex $v$, including its main process $P_v$.
\item The main process should be able to communicate (at least
  indirectly) in both ways with each helper process.
\item While communication links within one
  processor can be created freely, links between
  processes on different processors are only possible if there is a
  physical connection, that is a connection in the original
  peer-to-peer network; also, if there was a connection in the
  original network, then there should be one in the extension in order to
  preserve the network structure.
\item Lastly, to preserve symmetry, each automorphism of the original network
  must have an extension which maps all helper processes to the same
  processor as their corresponding main process.
\end{enumerate}

\begin{figure}[htb]
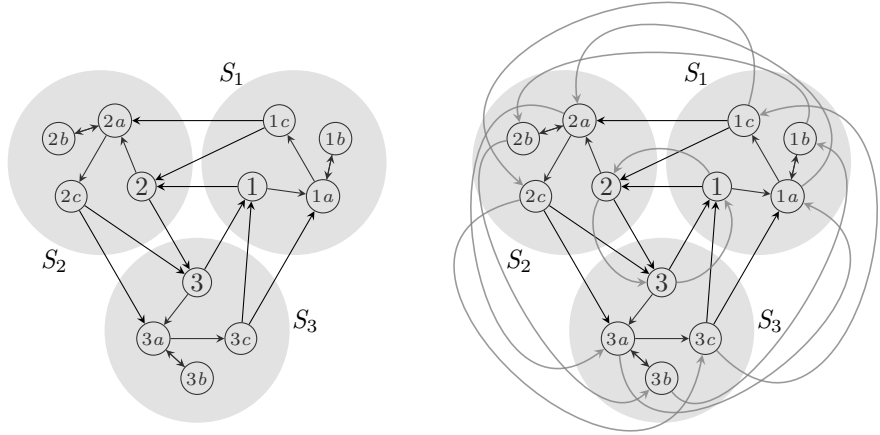

  \verboseornot{}{\vspace{-10mm}}
  \centering
  \subfloat[Symmetry-preserving extension of the network from
    \cref{fig:exampleGraph}\protect\subref{fig:exampleGraph:p2p}.]{
%    \hspace{-2mm}%
    \beginpgfgraphicnamed{graphic5_1}%
    \exampleNetworkAndExtension{x}{}%
    \endpgfgraphicnamed%
  }\hfill%
  \subfloat[Extended automorphism $\iota_\sigma$ as required by
  \cref{dfn:p2pext}.]{
    \beginpgfgraphicnamed{graphic5_2}%
    \exampleNetworkAndExtension{x}{x}%
    \endpgfgraphicnamed%
  }%
  \caption[Symmetry-preserving extension of the network from
    \cref{fig:exampleGraph}\protect\subref{fig:exampleGraph:p2p}.]{A
      symmetry-preserving extension (illustrating \cref{dfn:p2pext}).}
  \label{fig:p2pext}
  \verboseornot{}{\vspace{-1mm}}
\end{figure}

\begin{example}
  \label{ex:p2pext}
  See \cref{fig:p2pext} for an example of a symmetry-preserving
  extension. Note that condition (\ref{dfn:p2pext:interSv}) of
  \cref{dfn:p2pext} is liberal enough to allow helper processes to
  communicate directly with processes running on other processors, and
  indeed, e.g. $2c$ has a link to $3$. It also allows several
  communication links on one physical connection, reflected by the
  fact that there are three links connecting~$S_2$
  to~$S_3$.\verboseornot{ Furthermore, (\ref{dfn:p2pext:Svconn}) is
    satisfied in that the main processes are strongly connected with
    their helper processes, although, as e.g. with $2$ and $2c$,
    indirectly and through processes on other processors.}{}
\end{example}
We will need the following immediate fact later on.
\begin{fact}\label{fct:strongconn}
  As a direct consequence of \cref{dfn:p2p,dfn:p2pext}, any
  symmetry-preserving extension of a peer-to-peer network is strongly
  connected.
\end{fact}

\subsection{$G$-symmetry}
\label{sec:g-symmetry}

Corresponding to the intuition of processors with main and
helper processes, we weaken \cref{dfn:symmetricsystem}
such that only automorphisms are considered which keep the set of main
processes invariant and map helper processes to the same processor
as their main process. There are cases\verboseornot{}{ (as in
\cref{fig:networkVSymm} later in this paper)} where the main
processor otherwise would be required to run the same program as some
helper process.

\begin{definition}[$G$-symmetry]
  \label{dfn:g-symmetricsystem}
  A system \PP whose communication graph $G'$ is a symmetry-preserving
  extension of some peer-to-peer network $G=(V,E)$ is called
  \emph{$G$-symmetric} if \cref{dfn:symmetricsystem} holds
  with respect to those automorphisms $\sigma\in\Sigma_{G'}$
  satisfying, for all $v\in V$,
  \verboseornot{\begin{enumerate}[(i)]}{\begin{inparaenum}[(i)]}
  \item $\sigma(V)=V$ and
  \item $\sigma(S_v)=S_{\sigma(v)}$.
  \verboseornot{\end{enumerate}}{\end{inparaenum}}
\end{definition}
\verboseorextremenot{
This is weaker than \cref{dfn:symmetricsystem}, since
there we require the condition to hold for all automorphisms.

\begin{example}
  \label{ex:g-symmetricsystem}
  To illustrate the impact of $G$-symmetry,
  \cref{fig:g-symmetricsystem} shows a network $G$ and an extension
  where symmetry relates all processes which each other.
  $G$-symmetry disregards the automorphism which causes this
  and considers only those which keep the set of main processes invariant,
  i.e. the nodes of the original network $G$, thus allowing them to behave
  differently from the helper processes.
}{}
\verboseornot{
  
  Note that the main
  processes do not have a direct connection in the extension,
  which is permitted by \cref{dfn:p2pext} although it will
  obviously make it impossible for them to synchronize.
}{}
\verboseorextremenot{
\end{example}

\begin{figure}[htb]
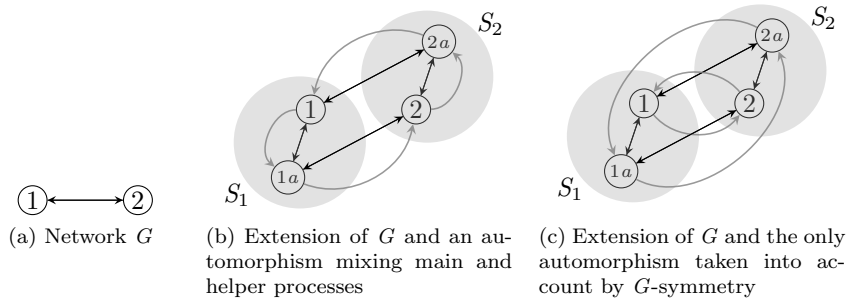

  \vspace{-5mm}
  \centering
  \subfloat[Network $G$]{
    \quad%
    \beginpgfgraphicnamed{graphic6}%
    \exampleNetworkVSymm{}{}%
    \endpgfgraphicnamed%
    \quad%
  }\quad%
  \subfloat[Extension of $G$ and an automorphism mixing main and helper
  processes]{
    \label{fig:g-symmetricsystem:mix}
%    \quad%
    \beginpgfgraphicnamed{graphic7}%
    \exampleNetworkVSymm{x}{x}%
    \endpgfgraphicnamed%
%    \quad%
  }\quad%
  \subfloat[Extension of $G$ and the only automorphism taken into account by
  $G$-symmetry]{
    \label{fig:g-symmetricsystem:invar}
%    \quad%
    \beginpgfgraphicnamed{graphic8}%
    \exampleNetworkVSymm{x}{gsymm}%
    \endpgfgraphicnamed%
%    \quad%
  }
  \caption{A network $G$ and an extension which has an automorphism
  mixing main and helper processes, disregarded by $G$-symmetry.}
  \label{fig:g-symmetricsystem}
  \verboseornot{}{\vspace{-5mm}}
\end{figure}
}{}
\verboseornot{

Now that we have formalized peer-to-peer networks and the
symmetry-pre\-ser\-ving extensions which we want to allow, we are
ready to prove positive and negative results about feasibility of
pairwise synchronization.
}{}

\section{Results}
\label{sec:results}

\subsection{Positive Results}
\verboseornot{First, we state the intuition foreshadowed
 in \cref{sec:pairw-synchr}, namely that \CSPio does allow for
 symmetric pairwise synchronization in peer-to-peer networks.
}{}

\begin{theorem}
\label{thr:p2p-allows-cspio}
  Let $G=(V,E)$ be a peer-to-peer network. Then $G$ admits a symmetric
  system pairwise synchronizing $V$ in \CSPio.
\end{theorem}
\begin{proof}
  A system which at each vertex $v\in V$ runs the program shown
  \verboseornot{in \cref{fig:p2p-allows-cspio}}{below} is symmetric
  and pairwise synchronizes $V$. Each process simply waits for each
  other process in parallel to become ready to send or receive a
  dummy message, and exits once a message has been exchanged with
  each other process.
  \verboseornot{\qedhere}{}
\end{proof}
\verboseornot{
\begin{figure}[htb]
  \verboseornot{}{\vspace{-5mm}}
  \centering
}{\small}
  \begin{csp}
    \lForEach{$w\in V$}{$sync_w\gets\false$}\;
    $\kern-1.5ptW_{in}\gets\{w\in V|(w,v)\in E\}$\;
    $\kern-1.5ptW_{out}\gets\{w\in V|(v,w)\in E\}$\;
    \Do{\;
      \lAlt{$\choice_{w\in W_{in}}\neg sync_w\wedge\rcv{P_w}{x}$}{$sync_w\gets\true$}\;
      \lAlt{$\choice_{w\in W_{out}}\neg sync_w\wedge\snd{P_w}{0}$}{$sync_w\gets\true$}\;
    }\;
  \end{csp}
\verboseornot{
  \caption[The program run at each vertex in the proof of
    \cref{thr:p2p-allows-cspio}.]{The program run at each vertex
      $v\in V$ in the proof of \cref{thr:p2p-allows-cspio}.}
  \label{fig:p2p-allows-cspio}
  \verboseornot{}{\vspace{-5mm}}
\end{figure}
}{\qedhere\normalsize}

\verboseornot{As a second result, we show that by}{By} dropping the
topological symmetry requirement for peer-to-peer networks, under
certain conditions we \verboseornot{can achieve symmetric pairwise
  synchronizing systems even in}{get a positive result even for}
\CSPin.

\begin{theorem}
  \label{thr:not-wamoti-allows-cspin}
  Let $G=(V,E)$ be a network satisfying only the first two conditions of
  \cref{dfn:p2p}, i.e. $G$ is strongly connected and directly
  connected. If $G$ admits a symmetric electoral system and
  there is some vertex $v\in V$ such that $(v,a)\in E$ \emph{and} $(a,v)\in
  E$ for all $a\in V$, then $G$ admits a symmetric system
  pairwise synchronizing $V$ in \CSPin.
\end{theorem}
\verboseornot{\begin{proof}}{\begin{proof}[sketch]}
  First, the electoral system is run to determine a temporary leader~$v'$.
  When the election has terminated, $v'$ chooses a coordinator $v$ that
  is directly and in both directions connected to all other vertices,
  and broadcasts its name. Broadcasting can be done by choosing a spanning tree
  and transmitting the broadcast information together with the
  definition of the tree along the tree, as in the
  proof of~\cite[Theorem~2.3.1, Phase 2]{bouge_existence_1988} (the
  strong connectivity which is required there holds for $G$ by assumption).
  After termination of this phase, the other processes each send one
  message to $v$ and then wait to receive commands from $v$ according
  to which they perform direct communications with each other, while
  $v$ receives one message from each other process and
  uses the obtained order to send out the commands.
  \verboseornot{

   This can be achieved by
   running the following program at each process $P_c$, $c\in V$, after
   having elected the temporary leader $v'$:
   \begin{itemize}
   \item If $c=v'$, choose some $v\in V$ such that $(v,a)\in E$ and
     $(a,v)\in E$ for all $a\in V$, and broadcast the name $v$;
     otherwise obtain the broadcast name.
   \item If $c=v$:
     \begin{itemize}
     \item Receive exactly one message from each other process in some
       non-deterministic order and remember the order:
       \begin{csp}
         $W\gets V\setminus\{v\}$\;
         \lForEach{$w\in W$}{$order_w\gets-1$}\;
         $count\gets0$\;
         \Do{
           \Alt{$\choice_{w\in W}order_w=-1\wedge\rcv{P_w}{x}$}{
             $order_w\gets count$\;
             $count\gets count+1$\;
           }\;
         }\;
       \end{csp}
     \item Issue commands to the other processes according to the
       obtained order:
       \begin{csp}
         \ForEach{$a,b\in V\setminus\{v\},a\neq b$}{
           \If{
             \Alt{$order_a<order_b\wedge(a,b)\in E$}{
               $\snd{P_a}{\text{``contact $b$''}}$\;
               $\snd{P_b}{\text{``listen to $a$''}}$\;
             }\;
             \Or{$order_a\geq order_b\vee(a,b)\not\in E$}{
               $\snd{P_b}{\text{``contact $a$''}}$\;
               $\snd{P_a}{\text{``listen to $b$''}}$\;
             }\;
           }\;
         }\;
       \end{csp}
     \end{itemize}
     otherwise (i.e. $c\neq v$):
     \begin{itemize}
     \item Send dummy message to $P_v$:
       \begin{csp}
         $\snd{P_v}{0}$\;
       \end{csp}
     \item Execute the commands from $v$ until one message has been
       exchanged with each other process:
       \begin{csp}
         $num\gets\abs{V\setminus\{c,v\}}$\;
         \Do{
           \Alt{$num>0\wedge P_v?m$}{
             \If{
               \lAlt{$m=\text{\upshape ``contact $w$''}$}{$\snd{P_w}{0}$}\;
               \lOr{$m=\text{\upshape ``listen to $w$''}$}{$\rcv{P_w}{x}$}\;
             }\;
             $num\gets num-1$\;
           }\;
         }\;
       \end{csp}
       \qedhere
     \end{itemize}
   \end{itemize}
 }{\qedhere}
\end{proof}

\begin{example}
  See \cref{fig:positive} for an example of a network which admits a
  symmetric system pairwise synchronizing all its vertices in \CSPin.
  The fact that the network admits a symmetric electoral system can be
  established as
  for~\cite[Fig.~4]{bouge_existence_1988}\verboseornot{. There the
    property is used that $\{1,2\}$ and $\{3,4,5\}$ are invariant
    under the network's automorphism group and the associated
    processes can thus behave differently; this property is not
    affected by the edges we have added}{}
  (note that the edges between the lower nodes are only in one
  direction).
\end{example}

\begin{figure}[htb]
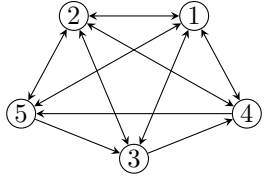

%  \verboseornot{}{\vspace{-5mm}}
  \centering
  \beginpgfgraphicnamed{graphic9}%
  \exampleAllowingSyncCSPin%
  \endpgfgraphicnamed%
  \caption[A network which admits a symmetric system pairwise
  synchronizing all its vertices in \CSPin.]{A network which by
    \cref{thr:not-wamoti-allows-cspin} admits a symmetric system
    pairwise synchronizing all its vertices in \CSPin.\verboseornot{
      Note that the connections between vertices $3$, $4$ and $5$ are
      only in one direction.}{}}
  \label{fig:positive}
%  \verboseornot{}{\vspace{-5mm}}
\end{figure}

This result could be generalized, e.g. by weakening the conditions on
$v$ and taking care that the commands will reach the nodes
at least indirectly. Since our main focus is the negative result, we will not
pursue this further.

\subsection{Negative Result}

In the following we will establish the main result saying that, even
if we extend a peer-to-peer network $G$ by helper processes (in a
symmetry-preserving way), it is not possible to obtain a network which
admits a $G$-symmetric system pairwise synchronizing the nodes of $G$
in \CSPin.

To this end, we derive a contradiction with
\cref{thr:bouge:3.3.2} by proving the following intermediate steps
(let $G$ denote a peer-to-peer network and $G'$ a symmetry-preserving
extension):
\begin{itemize}
\item \Cref{lm:sync_electoral}:
  If $G'$ admits a $G$-symmetric system pairwise
  synchronizing the nodes of $G$ in \CSPin, it admits a
  $G$-symmetric electoral system in \CSPin.
\item \Cref{lm:wamoti}:
  $G'$ has a non-trivial well-balanced automorphism
  taken into account by $G$-symmetry (i.e. satisfying
  the two conditions of \cref{dfn:g-symmetricsystem}).
\item \Cref{lm:eliminate_g}:
  We can extend $G'$ in such a way that
  there exists a non-trivial well-balanced automorphism (derived from
  the previous result), $G$-symmetry is reduced to symmetry,
  and admittance of an electoral system is preserved.
\end{itemize}

\begin{lemma}
  \label{lm:sync_electoral}
  If some symmetry-preserving extension of a peer-to-peer
  network $G=(V,E)$ admits a $G$-symmetric system
  pairwise synchronizing $V$ in \CSPin, then it admits a
  $G$-symmetric electoral system in \CSPin.
\end{lemma}

\begin{proof}
  The following steps describe the desired electoral system
  (using the fact that under $G$-symmetry processes of nodes $\in V$
  may behave differently from those of nodes $\not\in V$):
  \begin{itemize}
  \item All processes run the assumed $G$-symmetric pairwise
    synchronization program, with the following modification for the
    processes in $\PP:=\{P_v|v\in V\}$ (intuitively this can be seen
    as a kind of knockout tournament, similar
    to the proof of~\cite[Theorem~4.1.2, Phase 1]{bouge_existence_1988}):
    \begin{itemize}
    \item Each of these processes has \verboseornot{an additional}{a}
      local variable \var{winning} initialized to \val{true}.
    \item After each communication statement with some other
      $P\in\PP$, insert a second communication statement with $P$ in
      the same direction:
      \begin{itemize}
      \item If it was a ``send'' statement, send the value of
        \var{winning}.
      \item If it was a ``receive'' statement, receive a
        Boolean value, and if the received value is \val{true}, set
        \var{winning} to \val{false}.
      \end{itemize}
    \end{itemize}
    Note that, since the program pairwise synchronizes
    $V$, each pair of processes associated to vertices in $V$ has had
    a direct communication at the end of execution, and thus
    there is exactly one process in the whole system which has a local variable
    \var{winning} containing \val{true}.
  \item After the synchronization program terminates the processes
    check their local variable \val{winning}. The unique process that
    still has value \val{true} declares itself the leader and
  broadcasts its name; all processes set their variable \var{leader}
  accordingly. As in the proof of \cref{thr:not-wamoti-allows-cspin},
    broadcasting can be done using a spanning tree.
    The required strong connectivity is guaranteed by \cref{fct:strongconn}.
    \qedhere
  \end{itemize}
\end{proof}

\begin{lemma}\label{lm:wamoti}
  For any symmetry-preserving extension $G'=(V',E')$ of a peer-to-peer
  network $G=(V,E)$, there is $\sigma'\in\wamoti{G'}$ such that
  $\sigma'(V)=V$ and $\sigma'(S_u)=S_{\sigma'(u)}$ for all $u\in V$.
\end{lemma}
\begin{proof}
% I think the proof can be simplified by taking as
% sigma' sigma for vertexes in V and sigma^k for other vertexes, where
% k is the ratio between the size of the orbit of the vertex and the
% size of the orbit of vertexes in V
  Take an arbitrary $\sigma\in\wamoti G$ (exists by \cref{dfn:p2p})
  and let $\iota$, to save indices, denote the $\iota_\sigma$ required
  by \cref{dfn:p2pext}. If $\iota\in\wamoti{G'}$ we are done;
  \verboseornot{
    otherwise we can construct a suitable $\sigma'$ from $\iota$ by
    ``slicing'' orbits of $\iota$ which are larger than the
    period of $\sigma$ into orbits of that size. See
    \cref{ex:slicingOrbits} for an illustration of the following
    proof.
  }{
    otherwise we construct a suitable $\sigma'$ (\cref{ex:slicingOrbits}
    illustrates this proof).
  }

  Let $p$ denote the period of $\sigma$ and pick
  an arbitrary $v\in V$. For simplicity, we assume that $\sigma$ has only
  one orbit; if it has several, the proof extends straightforwardly
  by picking one $v$ from each orbit\verboseornot{ and proceeding with
    them}{} in parallel.
  
  For all $u\in S_v$ let $p_{u}:=\abs{O_u^\iota}$
  and note that for all $t\in O_u^\iota$ we have $p_t=p_u$, and $p_{u}\geq p$ since
  $\iota$ maps each $S_v$ to $S_{\sigma(v)}$ and these sets
  are pairwise disjoint.
  We define $\sigma':V'\to V'$ \verboseornot{as follows:}{and then prove 
    the claims.}
  \begin{equation*}
    \sigma'(u):=
    \begin{cases}
      \iota^{p_u-p+1}(u) &\text{if $u\in S_v$}\\
      \iota(u) &\text{otherwise.}
    \end{cases}
  \end{equation*}
  \verboseornot{Now we can show that}{}
%TODO explain the following verbosely
  \begin{itemize}
  \item $\sigma'(V)=V$, $\sigma'\neq\id$: Follows from
    $\iota\restriction_V=\sigma$ and $p_v=p$ and thus
    $\sigma'\restriction_V=\sigma$ (where $f\restriction_X$ denotes
    the restriction of a function $f$ to the domain $X$)
  \item $\sigma'\in\Sigma_{G'}$: With (\ref{dfn:p2pext:iotaS}) from
    \cref{dfn:p2pext} we obtain that, for $u\in S_v$, $p_u$ must be a
    multiple of
    $p$, and $\sigma'(O_u^\iota\cap S_v)=\iota(O_u^\iota\cap S_v)$,
    thus $\sigma'$ is a permutation of $V'$ since $\iota$ is
    one. Furthermore, for $t,u\in
    S_v$, we have $\iota^{p_t(p_u-1)}(t)=t$ and
    $\iota^{p_u(p_t-1)}(u)=u$ and therefore\verboseornot{}{ $\sigma'$
      also inherits edge-preservation from $\iota$ by}
    \begin{align*}
%       (\sigma'(t),\sigma'(u))&=(\sigma'(\iota^{p_t(p_u-1)}(t)),\sigma'(\iota^{p_u(p_t-1)}(u)))\\
%       &=(\iota^{p_tp_u-p+1}(t),\iota^{p_tp_u-p+1}(u))\enspace,
      (\sigma'(t),\sigma'(u))&=(\iota^{p_t-p+1}(t),\iota^{p_u-p+1}(u))
      \verboseornot{\\&}{}
      =(\iota^{p_tp_u-p+1}(t),\iota^{p_tp_u-p+1}(u))\enspace\verboseornot{,}{.}
    \end{align*}
    \verboseornot{thus $\sigma'$ also inherits edge-preservation from $\iota$.}{}
  \item $\sigma'(S_u)=S_{\sigma'(u)}$, $\sigma'$ well-balanced: The
    above-mentioned fact that for all $u\in S_v$ we have
    $\sigma'(O_u^\iota\cap S_v)=\iota(O_u^\iota\cap S_v)$, together with
    (\ref{dfn:p2pext:iotaS}) from \cref{dfn:p2pext} implies that
    also $\sigma'(S_u)=S_{\sigma(u)}$ for all $u\in
    V$. For all $v'\in V'$, well-balancedness of $\sigma$ and
    disjointness of the $S_u$ imply that $\sigma'^q(v')\neq
    v'$ for $0<q<p$. On the other hand, since each orbit of $\sigma$
    has size $p$ and contains exactly one element from $S_v$ (namely
    $v$), we have that
    \begin{align*}
      \sigma'^p(v')&=\iota^{(p_u-p+1)+(p-1)}(v') &\text{for some $u\in
        O_{v'}^\iota$}\\
      &=\iota^{p_u}(v')=\iota^{p_{v'}}(v')=v'\enspace.\tag*{\qedhere}
    \end{align*}
  \end{itemize}
\end{proof}

\begin{example}
  \label{ex:slicingOrbits}
  \verboseornot{Consider the extended peer-to-peer network $G'$ shown in
    \cref{fig:slicingOrbits}\subref{fig:slicingOrbits:iota} with
    automorphism $\iota_\sigma$ as required by \cref{dfn:p2pext}. We
    illustrate the construction of $\sigma'$
    given in the proof of \cref{lm:wamoti}.

    We
  }{In \cref{fig:slicingOrbits}\subref{fig:slicingOrbits:iota}, we}
  have $p=2$ (the period of
  $\sigma=\iota_\sigma\restriction_{\{1,2\}}$),
  and we pick vertex $v=2$. For the elements of $S_2$,
  we obtain $p_2=p=2$ and $p_{2a}=p_{2b}=p_{2c}=6$\verboseornot{ since, e.g.,
  $O_{2a}^{\iota_\sigma}=\{2a,1a,2c,1b,2b,1c\}$}{}. Thus $\sigma'$ is
  defined as follows:
  \begin{equation*}
    \sigma'(u)=
    \begin{cases}
      \iota(u) &\text{if $u=2$}\\
      \iota^5(u) &\text{if $u\in S_2\setminus\{2\}$}\\
      \iota(u) &\text{if $u\in S_1$\enspace.}
    \end{cases}
  \end{equation*}
  \verboseornot{This $\sigma'$ is depicted in
    \cref{fig:slicingOrbits}\subref{fig:slicingOrbits:sigma}.
    All orbits have the same cardinality, namely~$2$, and the remaining
    claims of \cref{lm:wamoti} are also satisfied.
  }{This $\sigma'$, depicted in
    \cref{fig:slicingOrbits}\subref{fig:slicingOrbits:sigma},
    satisfies the claims of \cref{lm:wamoti}.
  }
\end{example}

\begin{figure}[htb]
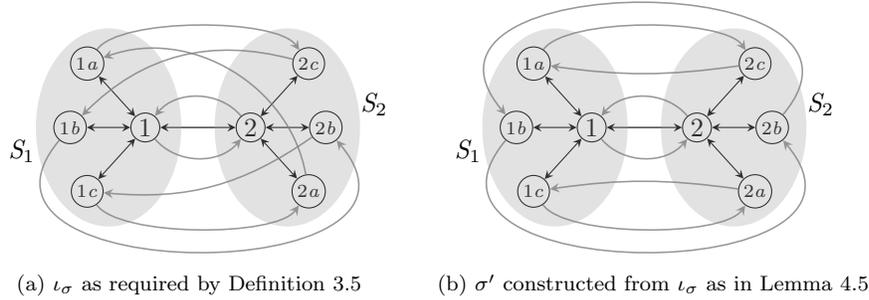

  \verboseornot{}{\vspace{-4mm}}
  \centering
  \subfloat[$\iota_\sigma$ as required by \cref{dfn:p2pext}]{
    \label{fig:slicingOrbits:iota}
    \beginpgfgraphicnamed{graphic10}
    \exampleSlicingOrbits{}
    \endpgfgraphicnamed
  }\verboseornot{}{\quad}%
  \subfloat[$\sigma'$ constructed from $\iota_\sigma$ as in
  \cref{lm:wamoti}]{
    \label{fig:slicingOrbits:sigma}
    \verboseornot{}{\quad}
    \beginpgfgraphicnamed{graphic11}
    \exampleSlicingOrbits{x}
    \endpgfgraphicnamed
    \verboseornot{}{\quad}
    \quad
  }
%  \verboseornot{}{\vspace{-1mm}}
  \caption{An extended peer-to-peer network $G'$ illustrating
    \cref{lm:wamoti}.}
  \label{fig:slicingOrbits}
  \verboseornot{}{\vspace{-5mm}}
\end{figure}

\begin{lemma}\label{lm:eliminate_g}
  Any symmetry-preserving extension $G'=(V',E')$ of a
  peer-to-peer network $G=(V,E)$ can be extended to a network
  $H$ such that
  \verboseorextremenot{\begin{enumerate}[(i)]}{\begin{inparaenum}[(i)]}
  \item $\wamoti{H}\neq\emptyset$, and
  \item if $G'$ admits a $G$-symmetric electoral system in \CSPin,\verboseorextremenot{\\}{}
    then $H$ admits a symmetric electoral system in \CSPin.
  \verboseorextremenot{\end{enumerate}}{\end{inparaenum}}
\end{lemma}
\begin{proof}
  The idea is to add an ``identifying structure'' to all elements of
  $V$, which forces all automorphisms to keep $V$ invariant and map
  the $S_v$ to each other correspondingly (see
  \cref{fig:networkVSymm}).
  Formally, let $K=\abs{V'}$ and, denoting
  the inserted vertices by $i_{.,.}$, for each $v\in V$ let
  \verboseornot{}{$I_v:=\bigcup_{k=1}^K\{i_{v,k}\}$ and}
  \begin{align*}
    \verboseornot{I_v &:=\bigcup_{k=1}^K\{i_{v,k}\}\\}{}
    E_v &:=\{(v,i_{v,1})\}\cup
    \bigcup_{k=1}^{K-1}\{(i_{v,k},i_{v,k+1}),(i_{v,k+1},v)\}\cup
    \bigcup_{w\in S_v}\{(i_{v,K},w)\}\enspace,
    \verboseornot{\\\intertext{and let}
    H &:=
    \Bigl(V'\cup\bigcup_{v\in V}I_v,E'\cup\bigcup_{v\in V}E_v\Bigr)\enspace.}{}
  \end{align*}
  \verboseornot{}{and let $H :=
    \Bigl(V'\cup\bigcup_{v\in V}I_v,E'\cup\bigcup_{v\in V}E_v\Bigr)$.}
  Now we can prove the two claims.
  \begin{enumerate}[(i)]
  \item Let $\sigma\in\wamoti{G'}$ with $\sigma(V)=V$ and
    $\sigma(S_v)=S_{\sigma(v)}$ for all $v\in V$ (such a $\sigma$
    exists by \cref{lm:wamoti}), then
    \verboseornot{\begin{equation*}}{$}
      \sigma\cup\bigcup_{v\in
        V}\bigcup_{k=1}^K\{i_{v,k}\mapsto i_{\sigma(v),k}\}\in\wamoti{H}
    \verboseornot{\enspace.\end{equation*}}{$.}
  \item $H$ is still a
    symmetry-preserving extension of $G$ via (straightforward)
    extensions of the $S_v$.
    The discriminating construction
    \verboseornot{
      (notably the fact that the vertices from $V$ now are guaranteed to
      have more edges than any vertex not in $V$, but still the same
      number with respect to each other)}{}
    has the effect that
    $\Sigma_H$ consists only of extensions, as above, of
    those $\sigma\in\Sigma_{G'}$ for which $\sigma(V)=V$ and
    $\sigma(S_v)=S_{\sigma(v)}$ for all $v\in V$. Thus, any
    $G$-symmetric system with communication graph~$H$ is
    a symmetric system with communication graph~$H$.

    Additionally, the set of all $i_{v,k}$ is invariant under
    $\Sigma_H$ due to the distinctive structure of the $I_v$,
%     any $\sigma\in\Sigma_H$ must map the $I_v$ to each other due to their
%     distinctive connection structure,
    thus the
    associated processes are allowed to differ from those of the remaining
    vertices.
    A symmetric electoral system in \CSPin can thus be
    obtained by running the original $G$-symmetric electoral system on all
    members of $G'$ and having each $v\in V$ inform $i_{v,1}$ about
    the leader, while all $i_{v,k}$ simply wait for and transmit the
    leader information.
    \qedhere
  \end{enumerate}
\end{proof}
\begin{figure}[htb]
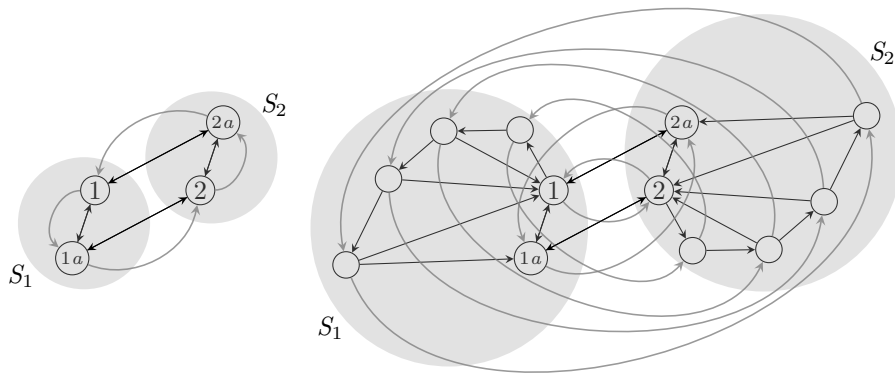

%  \verboseornot{}{\vspace{-5mm}}
  \centering
  \beginpgfgraphicnamed{graphic12}%
  \exampleNetworkVSymm{x}{x}%
  \verboseornot{}{\qquad}
  \exampleNetworkVSymm{idstr}{x}%
  \endpgfgraphicnamed%
  \caption[\verboseornot{The network from
    \cref{fig:g-symmetricsystem}}{A network} with an
  automorphism disregarded by $G$-symmetry, and the
  extension given in \cref{lm:eliminate_g}.]{\verboseornot{The
      network from \cref{fig:g-symmetricsystem}, shown}{A network}
    with an automorphism disregarded by $G$-symmetry, and the
    extension given in \cref{lm:eliminate_g} invalidating
    automorphisms of this kind shown with the only remaining
    automorphism.}
  \label{fig:networkVSymm}
%  \verboseornot{}{\vspace{-5mm}}
\end{figure}
\verboseornot{
Now we have gathered all prerequisites to prove our main
result.
}{}

\begin{theorem}\label{thr:main}
  There is no symmetry-preserving extension of any peer-to-peer
  network $G=(V,E)$ that admits a $G$-symmetric system
  pairwise synchronizing $V$ in \CSPin.
\end{theorem}

\begin{proof}
  Assume there is such a symmetry-preserving extension $G'$. Then by
  \cref{lm:sync_electoral} it also admits a $G$-symmetric electoral system
  in \CSPin. According to \cref{lm:eliminate_g}, there is then a
  network $H$ with $\wamoti{H}\neq\emptyset$ that admits a symmetric
  electoral system in \CSPin. This is a contradiction to
  \cref{thr:bouge:3.3.2}.
  \qedhere
\end{proof}

\section{Conclusions}
\label{sec:conclusions}

We have provided a formal definition of peer-to-peer networks and adapted
a semantic notion of symmetry for process systems communicating via
such networks. In this context, we have defined and investigated the
existence of pairwise synchronizing systems, which are directly
useful because they achieve synchronization, but also because they
create common knowledge between processes. Focusing on two dialects of
the \CSP calculus, we have proved the existence of such
systems in \CSPio, as well as the impossibility of implementing them
in \CSPin, even allowing additional helper processes like
buffers. We have also mentioned a recent extension to JCSP to show that,
while \CSPin is less complex and most commonly implemented,
implementations of \CSPio are feasible and do exist.

A way to circumvent our impossibility result is to remove some
requirements. For example, we have \verboseornot{provided}{sketched}
a construction for non-symmetric systems in \CSPin. In general, if we
give up the symmetry requirement, \CSPio can be implemented in
\CSPin~\cite[p.~197]{bouge_existence_1988}.

\verboseornot{
Another way is to tweak the definition or the assumptions about common
knowledge. Various possibilities are given in~\cite{halpern_knowledge_1990}.
By following the eager protocol proposed there,
common knowledge can eventually be
attained, but the trade-off is an indefinite time span during
which the knowledge states of the processes are inconsistent. This may
not be an option, especially in systems which have to be able to
act sensibly and rationally at any time.
Alternatively, if messages are guaranteed to be delivered exactly
after some fixed amount of time, common knowledge can also be
achieved, but this may not be realistic in actual systems.
Finally, possibilities to approximate common
knowledge are described. Approximate common knowledge or finite mutual
knowledge}{Another way is to weaken the notion of common knowledge or
approximate it~\cite{halpern_knowledge_1990}, which}
may suffice in settings where the impact decreases
significantly as the depth of mutual knowledge
increases, see e.g.~\cite{weinstein_impact_2007}.

However, if one is interested in symmetric systems and exact common
knowledge, as in the game-theoretical settings described in
\cref{sec:motivation}, then our results show that \CSPio is a
suitable formalism, while \CSPin is insufficient. Already in the
introducing paper~\cite{hoare_communicating_1978}, the exclusion of
output guards from \CSP was recognized as reducing expressivity and
being programmatically inconvenient, and soon it was deemed
technically not
justified~\verboseornot{\cite{bernstein_output_1980,buckley_effective_1983,filman_coordinated_1984}}{\cite{bernstein_output_1980,buckley_effective_1983}}
and removed in later versions of \CSP~\cite[p.~227]{hoare_communicating_1985}.

Some existing proposals for implementations of input and output guards
and synchronous communication could be criticized for simply shifting
the problems to a lower level, notably for not being symmetric
themselves or for not even being strictly synchronous in real systems
due to temporal imprecision~\cite{halpern_knowledge_1990}.

However, it is often useful to abstract away from implementation issues on the
high level of a process calculus or a programming
language (see e.g.~\cite[Section~10]{kurki-suonio_towards_1986}).
For these reasons, we view our setting as an argument for implementing \CSPio rather than \CSPin.

\section*{Acknowledgments}
I would like to thank my supervisor Krzysztof Apt for his support,
helpful comments and suggestions,
as well as four anonymous referees for their feedback which helped to
improve the paper.

This research was supported by a GLoRiClass fellowship
funded by the European Commission (Early Stage Research Training
Mono-Host Fellowship MEST-CT-2005-020841).

\bibliography{all,misc}

\end{document}

%% file: pgfexternal.tex
\long\def\beginpgfgraphicnamed#1#2\endpgfgraphicnamed{\includegraphics{#1}}

%% file: util.tex
\ifx\pdfoutput\undefined\else\fi

\newcommand{\crefformats}[7]{%
  \crefformat{#3}{##2\ifthenelse{\equal{#4}{}}{}{#4~}#1##1#2##3}
  \Crefformat{#3}{##2\ifthenelse{\equal{#5}{}}{}{#5~}#1##1#2##3}
  \crefrangeformat{#3}{\ifthenelse{\equal{#6}{}}{}{#6~}##3#1##1#2##4--##5#1##2#2##6}
  \Crefrangeformat{#3}{\ifthenelse{\equal{#7}{}}{}{#7~}##3#1##1#2##4--##5#1##2#2##6}
  \crefmultiformat{#3}{\ifthenelse{\equal{#6}{}}{}{#6~}##2#1##1#2##3}{, ##2#1##1#2##3}{ and~##2#1##1#2##3}
  \Crefmultiformat{#3}{\ifthenelse{\equal{#7}{}}{}{#7~}##2#1##1#2##3}{, ##2#1##1#2##3}{ and~##2#1##1#2##3}
  \crefrangemultiformat{#3}{\ifthenelse{\equal{#6}{}}{}{#6~}##2#1##1#2##3}{, ##2#1##1#2##3}{ and~##2#1##1#2##3}
  \Crefrangemultiformat{#3}{\ifthenelse{\equal{#7}{}}{}{#7~}##2#1##1#2##3}{, ##2#1##1#2##3}{ and~##2#1##1#2##3}
}

\makeatletter
\@ifclassloaded{llncs}{
  \crefformats{}{}{section}{Sect.}{Section}{Sects.}{Sections}
  \crefformats{}{}{subsection}{Sect.}{Section}{Sects.}{Sections}
  \crefformats{}{}{subsubsection}{Sect.}{Section}{Sects.}{Sections}
  \crefformats{}{}{chapter}{Chap.}{Chapter}{Chaps.}{Chapters}
  \crefformats{}{}{part}{Part}{Part}{Parts}{Parts}
  \crefformats{}{}{figure}{Fig.}{Figure}{Figs.}{Figures}
  \crefformats{}{}{table}{Table}{Table}{Tables}{Tables}
}{
   \crefformats{}{}{section}{Section}{Section}{Sections}{Sections}
   \crefformats{}{}{subsection}{Section}{Section}{Sections}{Sections}
   \crefformats{}{}{subsubsection}{Section}{Section}{Sections}{Sections}
   \crefformats{}{}{chapter}{Chapter}{Chapter}{Chapters}{Chapters}
   \crefformats{}{}{part}{Part}{Part}{Parts}{Parts}
   \crefformats{}{}{figure}{Figure}{Figure}{Figures}{Figures}
   \crefformats{}{}{table}{Table}{Table}{Tables}{Tables}
}
\makeatother

\newcommand{\marginlabel}[2]{%
  \mbox{}%
  \marginpar[\raggedleft\hspace{0pt}#1]{\raggedright\hspace{0pt}#2}%
}

\ifx\tikzstyle\undefined
\newcommand{\todoar}[2][]{\todo[#1]{#2}}
\else\tikzstyle{todoarrow}=[opacity=0.4,gray,-stealth]
\newcommand{\todoar}[2][]{%
  \marginlabel{\small #2}%\tikz[remember picture,overlay]\node(todoarrowstart){};}%
        {\tikz[remember picture,overlay,baseline=(todoarrowstart.220)]\node(todoarrowstart){};$\lhd$ \small #2}%
  \ifthenelse{\equal{#1}{}}{}{{\color{red}[}#1{\color{red}]}}%
  \tikz[remember picture,overlay]\node[inner sep=2pt](todoarrowend){};%
  \tikz[remember picture,overlay]\path(todoarrowstart)edge[todoarrow,out=190,in=-45](todoarrowend);%
}
\fi

\newcommand{\todo}[2][]{%
  \marginlabel{\small #2}{$\lhd$ \small #2}%
  \ifthenelse{\equal{#1}{}}{}{{\color{red}[}#1{\color{red}]}}%
}

\newcounter{autoexternalpgf}

%% file: mathutil.tex
\newcommand{\abs}[1]{\lvert#1\rvert}

\makeatletter
\@ifclassloaded{beamer}{}{%
  \newcommand{\newtheoremwithalias}[3]{%
    \newaliascnt{#1}{#2}
    \newtheorem{#1}[#1]{#3}
    \aliascntresetthe{#1}
  }

  \@ifclassloaded{llncs}{
    \if@envcntsame\errmessage{cleveref naming doesn't work because no aliascntrs used in llncs.cls}
    \else\if@envcntsect\newtheorem{fact}{Fact}[section]
    \else\newtheorem{fact}{Fact}
    \fi\fi
    \newcommand{\qedhere}{\qed}
    \crefformats{(}{)}{equation}{}{Equation}{}{Equations}
  }{
    \newtheorem{theorem}{Theorem}[section]
    \newtheoremwithalias{lemma}{theorem}{Lemma}
    \newtheoremwithalias{proposition}{theorem}{Proposition}
    \newtheoremwithalias{corollary}{theorem}{Corollary}
    \newtheoremwithalias{fact}{theorem}{Fact}
    \newtheoremwithalias{conjecture}{theorem}{Conjecture}
    \newtheoremwithalias{numberednote}{theorem}{Note}
    \newtheoremwithalias{numberedremark}{theorem}{Remark}

    \ifx\theoremstyle\undefined\else\theoremstyle{remark}\fi
    \newtheorem*{note}{Note}
    \newtheorem*{remark}{Remark}

    \ifx\theoremstyle\undefined\else\theoremstyle{definition}\fi
    \newtheoremwithalias{definition}{theorem}{Definition}
    \newtheoremwithalias{example}{theorem}{Example}

    \crefformats{(}{)}{equation}{Equation}{Equation}{Equations}{Equations}

  }

  \crefformats{}{}{theorem}{Theorem}{Theorem}{Theorems}{Theorems}
  \crefformats{}{}{lemma}{Lemma}{Lemma}{Lemmas}{Lemmas}
  \crefformats{}{}{proposition}{Proposition}{Proposition}{Propositions}{Propositions}
  \crefformats{}{}{corollary}{Corollary}{Corollary}{Corollaries}{Corollaries}
  \crefformats{}{}{fact}{Fact}{Fact}{Facts}{Facts}
  \crefformats{}{}{conjecture}{Conjecture}{Conjecture}{Conjectures}{Conjectures}
  \crefformats{}{}{numberednote}{Note}{Note}{Notes}{Notes}
  \crefformats{}{}{numberedremark}{Remark}{Remark}{Remarks}{Remarks}
  \crefformats{}{}{definition}{Definition}{Definition}{Definitions}{Definitions}
  \crefformats{}{}{example}{Example}{Example}{Examples}{Examples}
}
\makeatother

%% file: csp.tex
\usepackage{zed-csp,algorithm2e}

\newcommand{\snd}[2]{\ensuremath{#1\,!\,#2}}
\newcommand{\rcv}[2]{\ensuremath{#1\,?\,#2}}

\newenvironment{csp}{%
\newcommand{\choice}{\extchoice}
\renewcommand{\gets}{:=}
\SetKwSty{}%
\SetKwFor{While}{while}{do}{done}%
\newcommand{\DDo}[2]{\hspace{-2mm}{\bf$*$[\hspace{1.5mm}}##2 {\bf]}}%
\newcommand{\Do}[1]{\DDo{}{##1}}%
\renewcommand{\If}[1]{{\bf[\hspace{1.5mm}}##1 {\bf]}}%
\SetKwIF{Alt}{Or}{Else}{\kern-3pt}{$\rightarrow$}{$\extchoice$}{$\extchoice$ true $\rightarrow$}{}%
\SetKwFor{ForEach}{for each}{do}{done}%
\renewcommand{\Alt}[2]{\uAlt{##1}{##2}}%
\renewcommand{\Or}[2]{\uOr{##1}{##2}}%
\SetKwSwitch{Switch}{Case}{Other}{switch}{}{case}{else}{end}%
\SetKwComment{Cmt}{-- }{}%
\dontprintsemicolon%
\begin{algorithm}[H]}{\end{algorithm}}

\newcommand{\CSP}[1][]{\ensuremath{\mathit{C\kern-1ptS\kern-1ptP}_{\kern-1pt\mathit{#1}}}\xspace}

\newcommand{\CSPin}{\CSP[in]}
\newcommand{\CSPio}{\CSP[i\kern-1pt/\kern-1pto]}